\documentstyle[aps,prl,multicol,epsf]{revtex}

\renewcommand{\narrowtext}{\begin{multicols}{2} \global\columnwidth20.5pc}
\renewcommand{\widetext}{\end{multicols} \global\columnwidth42.5pc}
\renewcommand{\v}[1]{{\bf #1}}

\newcommand{\s}{{\sigma}}

\newcommand{\rb}{{\bar{\rho}}}

\def\eqa{\begin{eqnarray}}
\def\eea{\end{eqnarray}}
\newcommand{\eq}{\begin{equation}}
\newcommand{\ee}{\end{equation}}
\newcommand{\nn}{\nonumber\\}
\newcommand{\Eq}[1]{Eq.~(\ref{#1})}
\newcommand{\p}{\partial}

\newcommand{\ra}{\rightarrow}

\newcommand{\cH}{ {\cal H} }

\newcommand{\cR}{ {\cal R} }
\newcommand{\cS}{ {\cal S} }

\begin{document}
\draft
\title{Transport in Inhomogeneous Strongly Correlated Systems}
\author{Darwin Chang}
\address{Lawrence Berkeley Laboratory,
University of California, Berkeley, CA 94720, USA.}
\address{Physics Department, National Tsing-Hua University, Hsinchu 30043, 
Taiwan, R.O.C.}

\author{Dung-Hai Lee}
\address{Department of Physics, University of California, Berkeley, CA 94720.}
\maketitle
\begin{abstract}
We consider an inhomogeneous strongly correlated system where external
disorder divides it into mesoscopic cells.
Strong inter-particle repulsion suppresses the quantum tunneling between
cells and open a wide temperature range for incoherent transport.
For a particular class of such inhomogeneous systems we find  the low-energy
transport to be free-fermion like when the constituent particles are
spinless fermions.
\end{abstract}


\narrowtext

A central problem in today's condensed matter physics is the low-energy properties of doped Mott insulators. Here
we use the phrase ``Mott insulating'' loosely to denote any interaction-driven insulating behavior. Under this
definition  an insulating Wigner crystal is a Mott insulator, and a disorder-free Bose insulator is also a Mott
insulator. Unlike band insulators, Mott insulators can become spatially inhomogeneous upon doping. By
inhomogeniety we mean the {\it segregation} or {\it clustering} of doped particle density. Thus puddles or stripes
of doped regions embedded in undoped background are examples of inhomogeneity.  A Wigner crystal is not
inhomogeneous because particles localize in the smallest unit hence no clustering is exhibited.

Recently there are a number of examples where doped Mott insulators exhibit the tendency toward spatial
inhomogeneity.  This includes the stripes in the cuprates\cite{kiv} and nano-scale domains in the colossal
magneto-resistive magnites\cite{dagotto}.  In addition, in system such as near-critical thin superconducting films
and quantum Hall liquid near the plateau transition there are evidences that classical percolation is at work over
a significant temperature range\cite{kapi}.  These observations motivate us to think about an inhomogeneous
structure where the system is divided into cells of doped, strongly correlated, coherent subsystems of
intermediate length scale separated by an insulating medium. At zero temperature conduction is possible only
through quantum tunneling between cells.  We believe that this structure is formed for a wide class of doped Mott
insulators.

In addition to the difficulty imposed by inhomogeneity, the doped subsystems are poorly understood themselves.
This is because they are strongly correlated liquids, and except for one space dimension they are not well
characterized. For example basic questions such as ``what is the nature of low-lying excitations (quasiparticle or
collective)? and what is the nature of low-temperature transport ?'' remain unanswered.

In this paper 
we shall assume that in each cell there exists a wide energy (time) range where the single-particle Greens
function shows the ``Luttinger liquid'' behavior $G(\v x t';\v x t)\sim |t'-t|^{-(1+\alpha)}$ (here $\alpha > 0$)
for $\v x$ lies along the boundary of the cell\cite{note1}. Under this condition quantum tunneling is suppressed
at low energy. As a result there is a significant temperature range where the thermal energy exceeds the
renormalized bandwidth due to quantum tunneling, i.e. the transport is incoherent and thermal in origin. It is
possible that this intermediate incoherent transport regime is responsible for the unusual conduction behaviors of
the low-density two-dimensional electron gas\cite{krav}, near-critical thin superconducting films\cite{kapi}, and
nearly-percolating quantum Hall states\cite{shahar}.

Two concrete examples where the above postulated behaviors do
occur are i) an one dimensional Luttinger liquid divided into segments by an array
of impurities, and ii) a two dimensional quantum Hall liquids near the plateau
transition. In the following we shall discuss i) in considerable details and
comments on the necessary changes required for ii).  
We wish to demonstrate that these systems, while strongly coupled and disordered,  show surprising simple low
energy transport properties.


In 1D interacting (we shall focus on repulsive interactions) fermions and
bosons form so-called ``Luttinger liquid''\cite{haldane}. Under suitable
conditions, i.e. when the Luttinger liquid parameter is less than certain
critical value, a single impurity can change the low-energy transport
property entirely. This is because the impurity back scatters so strongly
at low energies that it cuts the system into two disconnected
halves\cite{kf}.
This is a specific example where strong correlation helps the impurity
potential to turn the system into spatial inhomogeneous cells.

Below certain cutoff energy $E_c$ the action describing $N$ impurities situated at $x_1,...,x_N$ in a Luttinger
liquid spanning $[0,L]$ ($L\ra\infty$) is given by\cite{note} \eqa &&\cS = \cS_0 + \cS_i\nn &&\cS_0= { 1 \over 2
\kappa} \int dx dt [ {1 \over v} (\p_t \phi)^2 + v (\p_x \phi)^2 ]\nn &&\cS_i = - \sum_{j=1}^N V_j \int dt
\cos(2\pi({\phi(x_j)\over \sqrt{\pi}} + \rb x_j)). \label{act} \eea In \Eq{act} $\rb$ is the average particle
density, $\phi(x)/\sqrt{\pi}+\rb x$ is the total number of particles contained in $(0, x)$, $\kappa$ is the
Luttinger liquid parameter, $v$ is a velocity, $V_j > 0$ is the back-scattering amplitude due to the $j$th
impurity. (We note that \Eq{act} also describes the pinning of a 1D charge density wave by impurities.)

We first consider the simple periodic case where $x_{j+1}-x_j=s={L\over N+1}$ for $0\le j\le N$ and
$V_j=V,~\forall j$ (here  $x_0 \equiv 0$, $x_{N+1} \equiv L$).
For free fermions ($\kappa =1$) such periodic arrangement will cause the formation of energy bands, and depending
on the value of $\rb s$ the system can either be a band metal ($\rb s\ne$ integer) or a band insulator ($\rb s =$
integer).  In the presence of a repulsive interaction ($\kappa <1$) the back scattering becomes a relevant
perturbation\cite{kf}. Thus even if $V$ starts out weak it will become strong at low energies.
In the following
we shall assume a strong {\it bare}
repulsive $V$.


For strong $V$ we can replace $\cS_i$ in \Eq{act} by a boundary condition \eq \phi(x_j)/\sqrt{\pi}+ \rb x_j=
m_j,~~~1\le j\le N. \label{bc} \ee So long as $V<\infty$ quantum tunneling is possible hence $m_{j+1}-m_j$, i.e.
the number of particles contained in each segment $(x_j,x_{j+1})$, is not a good quantum number. To describe such
physics we supplement the boundary condition \Eq{bc} with the following {\it weak} ($t<<E_c$) tunneling action
\eq
\cS_t=-t\sum_{j=1}^N
\cos(\sqrt{\pi}(\theta(x_{j-})-\theta(x_{j+})). \label{tun}
\ee
In \Eq{tun} $\theta(x)$ is the conjugate field of $\phi(x)$
(i.e., $[\phi(x),\theta(x)]=i$), and $x_{j\pm}$ are the two points to the
immediate right/left of $x_j$.

Let us temporarily ignore the weak quantum tunneling, and focus on the
effects of the boundary condition (\Eq{bc}).  Without tunneling all $m_j$'s
are time independent, and we can eliminate them from the boundary
condition by introducing  $\tilde{\phi}(x)$ so that for $x\in (x_j,x_{j+1})$
\eq
\phi(x)=\phi_j+\frac{\phi_{j+1}-\phi_j}{s}(x-x_j)+\tilde{\phi}(x).
\ee
Here $\phi_j\equiv\phi(x_j)$, and $\phi_0=\phi_{n+1}\equiv 0$ (we assume
$\rb L$, the total number of particles in $(0,L)$, = integer). Under this
change of variables the boundary condition becomes
$\tilde{\phi}(x_j)=0~~~\forall j$,
and the action becomes
\eqa
S_0&&=\sum_{j=0}^n\int_0^{\beta} dt L_j\nn
L_j&&={v\over 2\kappa}\frac{(\phi_{j+1}-\phi_j)^2}{s}+ {1 \over
2\kappa}\int_{x_j}^{x_{j+1}} dx [ {1 \over v} (\p_t \tilde{\phi})^2 + v
(\p_x \tilde{\phi})^2 ].\nn&&
\eea
We note that
\eqa \phi_{j+1}-\phi_j=\sqrt{\pi}[n_j-\rb s] \eea
where $n_j=m_{j+1}-m_j$. The partition function is given by
\eq
Z=\prod_j \{\sum_{n_j}\int_{\tilde{\phi}_j=\tilde{\phi}_{j+1}=0}
D[\tilde{\phi}] ~e^{-[\beta V(n_j)+\int_0^{\beta}dt L_j]}\}.
\label{z}
\ee
Here
\eq
V(n)= \frac{U}{2}(n-\rb s)^2,
\label{ce}
\ee
with
%
$U=\pi v/(\kappa s)$.
Thus each Luttinger liquid segment is characterized by $n_j$ and
$\tilde{\phi}$. The former specifies the total number of particle contained
between $(x_j,x_{j+1})$ , and the latter describes the collective density
fluctuations. \Eq{ce} has the form of ``charging energy'' discussed in the
literature. For $\kappa < 1$ (which corresponds to repulsive interaction
among the electrons) the charging energy $U$ is larger than
\eq
\Delta_0=v/s
\ee
 - the finite-size energy gap of the density fluctuation.

The ground state energy associated with \Eq{z} is
\eq
E=E_0+\sum_j \{{U\over 2} F(\rb s)-\frac{\pi}{24 v s}\}.
\ee
In the above $E_0$ is the ground state energy of the ideal system (i.e. without the impurities), $-\pi/24 v s$ is
the zero-point (Casimir) energy due to quantum fluctuations in $\tilde{\phi}$, and $F(\rb s)$ is due to the
charging energy. It is a periodic function given by
\eq
F(y)\equiv
- \lim_{\alpha\ra\infty}{1\over\alpha}\ln\{\sum_{m=-\infty}^{\infty}
e^{-\alpha (m-y)^2}\}.
\ee

Now we address the effects of quantum tunneling (\Eq{tun}). It is known
from the study of the single impurity problem that for $\kappa<1$ single
particle tunneling is suppressed at low energies.
To be more specific, take $t$ in Eq. (2) as the tunneling amplitude defined
at a microscopic energy scale $E_c$.
As shown by Kane and Fisher\cite{kf} the renormalized tunneling
$t_{\rm{eff}}(E)$ obeys the following equation
\eq
\frac{dt_{\rm{eff}}}{dl}=(1-\frac{1}{\kappa})~t_{\rm{eff}}.\label{fl}
\ee
Here $l=\ln(E_c/E)$ is the logarithmic
energy rescaling factor. Integrating this equation we obtain
$t_{\rm{eff}}(E)=t (E/E_c)^{(\kappa^{-1}-1)}$,
which implies the decrease of the effective
tunneling as $E\ra 0$. The renormalization group equation for
$t_{\rm{eff}}$ in the presence of many impurities is
in general different from \Eq{fl}. However it is simple to show that so
long as $E>\Delta_0$ \Eq{fl} continues to
apply. In the following we shall use \Eq{fl} until $E\sim U >\Delta_0$.
We then construct an effective Hamiltonian using the renormalized parameters
to address the physics below the energy $U$.

Let us first concentrate at zero temperature. Consider an renormalization
group transformation with the energy rescaling factor $E_c/U$. After this
transformation
$t\ra t_{\rm{eff}}(U)=t\left(U/ E_c\right)^{(\kappa^{-1}-1)}$,
$U\ra U_{\rm{eff}}=E_c$, and $\Delta_0\ra \Delta_{\rm{eff}} =
E_c\kappa/\pi$. Since
\eq \cR\equiv\frac{t_{\rm{eff}}(U)}{\Delta_{\rm{eff}}}= {\pi\over
\kappa}\frac{t}{E_c}\left(\frac{U}{E_c}\right)^{(\kappa^{-1}-1)}<<1, \ee
quantum tunneling clearly is incapable of exciting the collective modes
within the individual cells.
Under this condition each Luttinger liquid segment $(x_j,x_{j+1})$ is
described by a single quantum number $n_j$ - the total number of particles
in that segment. Since $t_{\rm{eff}}\ra 0$ we only need to consider the two
$n_j$ that are closest to $\rb s$.
%
Assuming $\rb s=(k+1/2)+f$ where $-1/2<f\le 1/2$ and $k$ is an integer, the
two lowest energy charge states are $n_j=k$ and $n_j=k+1$.  The difference
between the renormalized charging energy of these two states is
\eq U_{\rm{Q}}={U_{\rm{eff}}\over 2}({1\over 2}+f)^2-{U_{\rm{eff}}\over 2}({1\over 2}-f)^2= E_c f. \ee
If we denote the state with $n_j=k$ as $S^z=-1/2$ and the state with $n_j=k+1$ as $S_z=1/2$ the effective
Hamiltonian is given by
\eq \cH = -t_{\rm{eff}}\sum_j(S_j^+S_{j+1}^-+ S_{j+1}^+S_{j}^-)+ U_{\rm{Q}}\sum_j S_j^z . \label{xy} \ee
Because the total particle number $(N+1)(k+1/2+f)$ is a good quantum number, we are constrained to look at the
$S^z_{\rm{tot}}=(N+1)f$ sector of \Eq{xy}.  In one space dimension \Eq{xy} is equivalent to the free fermion
model. Thus at low energy the periodic array of Luttinger segments has free fermion quasiparticles!  This is
somewhat a surprise because in the absence of impurities the system is a repulsive Luttinger liquid for which
quasiparticle is not well defined.  For $f< 1/2$ the flipped (i.e. $S_z=-1/2$) spins (or the fermions) coherently
moves with a bandwidth of order
$t_{eff}$. When 
$f=1/2$ each segment has an unique ground state $k+1$ which is separated by an energy gap $U_{\rm{Q}}>>t_{eff}$
from the two closest excited charge states $k$ and $k+2$. Under this condition the system is an insulator.

Let us now turn on a non-zero temperature $T$ satisfying $U<< k_BT<<E_c$.
By choosing $E_c/K_BT$ to be the energy rescaling factor, we obtain the
following ratio between the renormalized quantum tunneling energy
($t_{\rm{eff}}(k_BT)$) and the rescaled thermal energy ($E_c$)
\eq
\frac{t_{\rm{eff}}(k_BT)}{E_c} =
\frac{t}{E_c}\left(\frac{k_BT}{E_c}\right)^{(\kappa^{-1}-1)}<<1\label{much}.
\ee
Thus due to the strong correlation there exist a wide temperature range
$(U,E_c)$ in which the thermal energy exceeds the renormalized
bandwidth, implying the transport is incoherent.
\\

The above results can be generalized to the situation where the impurities do not form a regular array. Changing
$x_{j+1}-x_j=s_j$ by the amount of average inter-particle distance is sufficient to cause $\rb s$ to change by
unity. Under this condition \Eq{xy} is replaced by \eq \cH = -t_{\rm{eff}}\sum_j( S_j^+S_{j+1}^-+
S_{j+1}^+S_{j}^-) + \sum_j w_j S_j^z. \ee This is equivalent to spinless free fermions moving in the presence of
disorder potential $w_j$. Since the variance of $w_j$ is of order $U_{\rm{eff}} >> t_{\rm{eff}}$ the system is in
strongly localized regime. Here we explicitly see how strong electron correlation enhances particle localization.
\\

After seeing the above one dimensional example one naturally wonders whether there are higher dimension analog of
it.  A closely related example is given by the the quantum percolation picture\cite{cc,lwk} of the fractional
quantum Hall plateau transition. In this picture fractional quantum Hall droplets form upon a smooth potential.
Quantum tunneling between these droplets is responsible for the delocalization of quasipartices. For concreteness,
let us focus on approaching the $\s_{xy}=0$ (Hall insulator) $\ra \s_{xy}=e^2/3h$ (quantum Hall liquid) transition
from the $\s_{xy}=0$ side.  Physically we can think of droplets of the $\nu=1/3$ incompressible liquids grow on an
insulating background. Let us temporarily ignore the electron tunneling between these droplets. For a given liquid
droplet with perimeter length $s_j$ the ``charging energy'' (i.e. the increment of the potential energy when an
electron is added to the droplet) is $ U= 6 \pi v/s_j$ where $v$ is a velocity. The finite-size gap of the chiral
collective edge mode is a factor of three smaller, i.e. $\Delta= 2\pi v/s_j$. By tuning the magnetic field or
tuning the gate voltage applied to the two dimensional electron gas we can change the size of the Hall droplets
and hence $s_j$.

Now let us switch on the quantum tunneling. Since the intervening region between quantum Hall droplets is an
insulating background, only electrons can tunnel.  It is well known in that the electron tunneling between two
$\nu=1/3$ droplets gets weaker as the energy decreases\cite{wen}, and the flow equation is the same as \Eq{fl}
where $\kappa=1/3$.  As the critical point of the plateau transition is approached the Hall droplets gets bigger
and hence $U$ becomes smaller.  When $E_c/U>>1$ ($E_c$ is the average energy required to inject and remove an
electron from the interior of each Hall droplet) there exists a wide dynamic range for $t_{\rm{eff}}$ to flow
down. When the renormalized tunneling becomes much smaller than the finite-size gap of the edge collective mode,
each droplet is charaterized by a single quantum number -- the number of electrons it contains.  Since the
renormalized tunneling is very weak we only need to consider the two lowest-energy charge states.  The low-energy
effective Hamiltonian then read \eq \cH=-\sum_{ij}[t_{ij,{\rm eff}} c^+_i c_j + h.c ] + \sum_i w_{j} c^+_ic_i.
\label{trp} \ee In the above $i,j$ labels the Hall droplets and $w_{j}$ is the random charging energy.  An
important new feature of \Eq{trp} is that $t_{ij,\rm{eff}}$ are complex numbers with random phases\cite{cc}. If we
ignore the small tunneling term in \Eq{trp} altogether the electrons will be localized on the droplets.  In that
case the plateau transition occurs via classical percolation.  As the droplet size changes in the classical
percolation process $\rb s_j$ varies. Every time $\rb s_j$ crosses half-integer value two charge states $m_j$ and
$m_j+1$ become degenerate. If the $\rb s_j$ of two droplets crosses half integer at the same time the quantum
tunneling between them can not be ignored. In this way at any stage of the classical percolation there will be a
set of $t_{ij,\rm{eff}}$ remains ``active''. It is the common belief that these active tunneling will eventually
change the universality of the delocalization transition\cite{lwk}. It is important to point out that the ultimate
critical behavior of the $0\rightarrow 1/3$ plateau transition is described by \Eq{trp} which can be used to
describe the $0\rightarrow 1$ integer plateau transition\cite{klz}. The above discussion also suggests that when
the disorder potential is very smooth (which results in small quantum tunneling amplitude) the true critical
behavior of fractional quantum Hall plateau transition can be difficult to see. For a wide energy/temperature
range the system will exhibit classical percolation behavior.

The essential ingredients responsible for the physics  discussed above are
the following. 1) Spatially inhomogeneous structures consists of strongly
correlated liquid cells linked by quantum tunneling. 2) A Luttinger liquid
behavior of the single particle Greens function over a significant energy
range
\eq
<\psi(x,t')\psi^+(x,t)>\sim |t'-t|^{-(1+\alpha)}, ~~ x\in {\rm
cell~boundary},
\label{asp}
\ee
where $\alpha>0$.
Besides the examples provided above, condition 1) is probably realized in a large class of doped Mott insulator.
The condition 2) which implies the flow equation \Eq{fl} with $\kappa\ra 1/(1+\alpha)$ guarantees that tunneling
is suppressed at low energy. As long as condition \Eq{fl} is applicable over a wide range of energy, there should
exists a wide range of temperature in which incoherent thermal transport dominates. It is possible that this is a
mechanism by which the true coherent quantum transport properties can be hidden by incoherent thermal transport
behavior in many of the materials observed.

Acknowledgements: DHL is supported in part by NSF grant DMR 99-71503 and DC by National Science Council, Taiwan,
ROC.  DC wishes to thank Theory Group of LBL (Berkeley) and that of SLAC(Stanford) for hospitality.

\widetext


\begin{thebibliography}{99}



\bibitem{kiv}
V. J. Emery, S. A. Kivelson and J. M. Tranquad, Proc. Natl. Acad.
Sci. (USA) {\bf 96}, 8814 (1999).

\bibitem{dagotto}
 J. Burgy {\it et al}, Phys. Rev. Lett. {\bf 87}, 277202 (2001);E. Dagotto, T. Hotta and A. Moreo, Phys. Rept. {\bf 344}, 1 (2001).

\bibitem{kapi}
 E. Shimshoni and A. Auerbach,
Phys. Rev. B{\bf 55}, 9817-9823 (1997);
E. Shimshoni,  A. Auerbach  and A. Kapitulnik,
Phys. Rev. Lett. {\bf 80}, 3352 (1998);
N. Mason and A. Kapitulnik,
Phys. Rev. Lett. {\bf 82}, 5341 (1999);
Leonid P. Pryadko and  A. Auerbach,
Phys. Rev. Lett., {\bf 82} 1253-1256, (99).

\bibitem{note1} Here we only require $G$ to exhibit the Luttinger liquid behavior
below certain relatively long space and time scales. Above these crossover scales quasiparticle could be
well-defined. This short-distance(time) behavior becomes relevant if the spatial crossover scale exceeds the
linear dimension of the cell, and the time crossover scale exceeds the inverse finite size gap. This is much less
abitious than the conjecture of P.W. Anderson that Luttinger liquid behavior is realized in the thermaldynamic
limit for $d>1$.

\bibitem{krav} For a review see
E. Abrahams, S. V. Kravchenko and M. P. Sarachik,
Rev. Mod. Phys. 73, 251 (2001).

\bibitem{shahar}  D. Shahar {\it et al}, Solid State Commun., {\bf107}, 19 (1998).

\bibitem{haldane}
F. D. M. Haldane, J. Phys. {\bf C14}, 2585 (1981).

\bibitem{kf}
C.L. Kane and M.P.A. Fisher, Phys. Rev. Lett. {\bf 68}, 1220 (1992);
Phys. Rev. {\bf B46}, 7268 (1992);
Phys. Rev. {\bf B46}, 15233 (1992).

\bibitem{note}Because the following discussions are intended for both
strongly interaction fermions and bosons, we shall ignore the fermion
spin.




\bibitem{cc} J. T. Chalker and P. D. Coddington,
J. Phys. {\bf C21}, 2665 (1988).

\bibitem{lwk} D-H. Lee, Z. Wang and S. Kivelson,
 Phys. Rev. Lett. {\bf 70}, 4130 (1993).

\bibitem{wen} X. G. Wen,
Adv. Phys. {\bf 44}, 405 (1995).

\bibitem{klz} D-H. Lee, S. Kivelson and S-C. Zhang
 Phys. Rev. Lett. {\bf 68}, 2386 (1992)

\end{thebibliography}
\end{document}